\RequirePackage{lineno}
\documentclass[twocolumn,superscriptaddress,prd,aps,preprintnumbers,amsmath,amssymb,nofootinbib]{revtex4}
\usepackage{graphicx}
\usepackage{dcolumn}
\usepackage{bm}
\usepackage{bbm}
\usepackage[dvips]{color}
\usepackage{cancel}
\usepackage{latexsym}
\usepackage{here} 
\usepackage{amsmath}
\usepackage{mathrsfs}
\usepackage{amssymb}  
\usepackage{multirow}
\usepackage{lineno}
\evensidemargin=\oddsidemargin
\newcommand{\tr}{\mbox{tr}}
\newcommand{\fb}{fb$^{-1}$}

\newcommand{\pt}{$p_{\mathrm{T}}$}

\newcommand{\MET}{$E_{\mathrm{T}}^{\mathrm{miss}}$}

\begin{document}

\preprint{}
\preprint{}

\title{Drell-Yan Production of $Z'$ in the Three-Site Higgsless Model at
the LHC}

\author{Tomohiro Abe}
\affiliation{Department of Physics, Nagoya University, Nagoya 464-8602, Japan}

\author{Tatsuya Masubuchi}
\author{Shoji Asai}
\author{Junichi Tanaka}
\affiliation{
The University of Tokyo, International Center for
Elementary Particle Physics and Department of Physics, 7-3-1
Hongo, Bunkyo-ku, JP - Tokyo 113-0033, Japan}

\date{\today}

\begin{abstract}
In the Higgsless models, there are extra gauge bosons which keep
the perturbative unitarity of a longitudinally polarized gauge boson. 
The three-site Higgsless model is a minimal Higgsless model and
contains three extra gauge bosons, $W^{\prime \pm}$ and $Z'$.
In this paper,
we report the discovery potential of the $Z'$ gauge boson via Drell-Yan production with
$Z'$(mass=380, 500, 600~GeV)$\rightarrow WW \rightarrow \ell\nu qq$~($\ell=e$, $\mu$) at the LHC~($\sqrt{s}$=14~TeV).
\end{abstract}

\maketitle

\section{Introduction}
The standard model (SM) describes the phenomenology of elementary 
particles very well. 
Its predictions are consistent with many experimental results.
Spontaneous symmetry breaking (SSB) is an important concept in the SM,
and
the electroweak symmetry breaking (EWSB) is known to be spontaneously
broken. 
In the SM, the EWSB is triggered by a Higgs boson.
However,
the Higgs boson has not been discovered yet in any experiment. 
This fact means the origin of EWSB still remains a mystery.
A Large Hadron Collider experiment at CERN (LHC)~\cite{LHC} has started with a center-of-mass energy of $\sqrt{s}=7$~TeV
and
the LHC is expected to reveal the origin of the EWSB.
%
There are two possibilities, that is, scenarios with and without
the Higgs boson to describe the EWSB.
We focus on the latter scenario in this paper.
%
In case the Higgs boson does
not exist, there is no longer so-called naturalness problem.
However, in this case there are
problems in the unitarity of the longitudinal gauge boson scattering
\cite{Herrero:1998eq,Dicus:1992vj,Cornwall:1973tb,Cornwall:1974km,Lee:1977yc,Lee:1977eg,Chanowitz:1985hj}
and
in the consistency with electroweak precision tests.


During the past decade, models with extra dimensions have been studied as a new
paradigm. It has brought a solution of the gauge hierarchy, a solution
of the Yukawa hierarchy, many new particles called Kaluza-Klein (KK)
particles and dark-matter candidates (with the symmetry called KK parity). 
The Higgsless model
\cite{Csaki:2003dt,Csaki:2003zu,Nomura:2003du,Barbieri:2003pr,Csaki:2003sh,Davoudiasl:2003me,Burdman:2003ya,Foadi:2003xa,Hirn:2004ze,Casalbuoni:2004id,Chivukula:2004pk,Csaki:2005vy,Csaki:2004sz}
is one of the extra-dimensional models. It does not
contain any physical scalar field. The EWSB is triggered by boundary
conditions for an extra-dimensional direction.

The Higgsless model can
keep a perturbativity,
and can be translated into a model in four dimensions
by the discretization of the extra-dimensional direction.
This translation is known as the deconstruction
\cite{Arkanihamed:2001ca,Hill:2000mu}.
The deconstruction allows us to interpret a gauge symmetry in
extra dimension as a direct product of infinite number of gauge
symmetries.
Hence a model with a direct product of finite number of gauge
symmetries can be
regarded as a low energy effective theory, where one of the ultraviolet
(UV) completions is described with
an extra-dimensional model.
Such models can be constructed by a bottom-up approach using non-linear
sigma models. 
The structure 
of their gauge sector
is based on a generalized hidden local symmetry \cite{Georgi:1989gp,Georgi:1989xy,Bando:1984ej,Bando:1985rf,Bando:1987ym,Bando:1987br,Harada:2003jx}.

The three-site Higgsless model \cite{SekharChivukula:2006cg}
is a minimal deconstructed Higgsless model
and contains three extra relatively heavy gauge bosons, $W^{\prime\pm}$ and $Z'$ as explained later.
The electroweak gauge symmetry of this model is $SU(2) \times SU(2) \times U(1)$,
which is broken down to $U(1)_{QED}$.
Hence this model is a low energy effective theory of the Higgsless model
in extra dimension and other UV complete models.
Through studies of this model,
we can find validity of other similar models at a time.
This is an advantage of this model, 
and 
it is important to find phenomenological constraints on this model.

Phenomenological constraints on this model are compatible with current
experiments \cite{SekharChivukula:2006cg,
Abe:2008hb,Abe:2009ni}.
Hence the LHC should be the most powerful experiment to test this model. 
Because we have already know the precise bounds on parameters in this
model,
we can accurately predict physics and signals at the LHC.
There are papers on the three-site Higgsless model for the LHC~\cite{Ohl:2008ri, 
Speckner:2010zi,
Han:2009em,Asano:2010ii,
Han:2009qr,Bian:2009kf,He:2007ge}, 
and
many of them are based on the so-called parton level analysis.
The parton level analysis ignores the effects of hadronizations and detector responses,
which are needed to get more realistic prospects.
Results, for example, the requirement of integrated luminosity
needed for new particle discovery, without considering such effects might be optimistic.
Therefore hadronizations and detector simulations are performed for the results shown in this paper.

According to the parton level analysis,
Drell-Yan (DY) production process of the $W^{\prime \pm}$ and $Z'$ bosons
is the most promising channel for their discovery.
Less integrated luminosity is required for the discovery of heavy gauge bosons 
through the DY process than others.
As explained later,
the coupling among the $Z'$ and fermions, $g_{Z'ff}$,
is stronger than 
the coupling among the $W'$ and fermions, $g_{W'ff}$ and
a parameter dependence of $g_{Z'ff}$ is more moderate compared with $g_{W'ff}$.
Therefore we focus on the DY production of $Z'$ in this paper, 
and study it with a way beyond the parton level analysis.

This paper is organized as follows.
In section \ref{sec:3site-review}, 
we review the three-site Higgsless model briefly.
%
%
In section \ref{sec:analysis},
we perform feasibility studies with the experimental condition of the ATLAS experiment~\cite{ATLASDetector} for some signal points. 
Section \ref{sec:summary} 
is devoted for summary and discussion.
\section{The three-site Higgsless model}
\label{sec:3site-review}
In this section we review 
the three-site Higgsless model
briefly.
This is a minimal Higgsless model, and its electroweak gauge symmetry is
$SU(2)_0 \times SU(2)_1 \times U(1)_2$. The electroweak symmetry breaking is
described using the Hidden local symmetry language, or non-linear sigma
fields.
We explain gauge and fermion sectors with its Lagrangian
and physical features for each sector.

\subsection{Gauge sector}
\label{3site-gauge-sec}
The gauge sector of this model is written as
\begin{eqnarray}
  {\cal L}_{\mathrm{gauge}}
&=&
-\frac{1}{4} G_{\mu\nu}^a G^{a\mu\nu} 
\label{eq:3site-gluon}
\\
&&
-\frac{1}{4}\sum_{i=0,1} W_{i\mu\nu}^a W_i^{a\mu\nu} 
           -\frac{1}{4}B_{\mu\nu} B^{\mu\nu} 
\nonumber
\\
&&
           +\sum_{i=1,2}\frac{f_i^2}{4}\tr\bigl[
             (D_\mu U_i)^\dagger (D^\mu U_i)
            \bigr] .
\label{eq:3site-gauge-sec}
\end{eqnarray}
Eq.~(\ref{eq:3site-gluon}) is a gluon sector, 
and Eq.~(\ref{eq:3site-gauge-sec}) are a electroweak sector.
$W_{0 \mu}$, $W_{1 \mu}$ and $B_{\mu}$ are gauge fields of $SU(2)_0$,
$SU(2)_1$ and $U(1)_2$, respectively. Their gauge couplings are $g_0$,
$g_1$ and $g_2$, respectively. 
$U_{i}$ are would-be Nambu-Goldstone bosons in non-linear sigma
representation\footnote{$\tau^a$ is Pauli matrices.}, 
\begin{eqnarray*}
  U_i &=& \exp \left(i\frac{\pi^a_i \tau^a}{f_i}\right),
\end{eqnarray*}
and their covariant derivatives are following;
\begin{eqnarray*}
D_\mu U_1 &=& \partial_\mu U_1 
              + i g_0 \dfrac{\tau^a}{2} W^a_{0\mu} U_1
              - i g_1 U_1 \dfrac{\tau^a}{2} W^a_{1\mu} \\
  D_\mu U_2 &=& \partial_\mu U_2 
              + i g_1 \dfrac{\tau^a}{2} W^a_{1\mu} U_2
              - i g_2 U_2 \dfrac{\tau^3}{2} B_{\mu}  .
\end{eqnarray*}
Notice that $U(1)_2$ gauge symmetry acts on $U_2$ as a ``partially gauged
$SU(2)_2$''.
All fields in the above Lagrangian are not mass eigenstate but gauge
eigenstate except gluon field. 
Mass matrices for gauge bosons can be derived from Eq.~(\ref{eq:3site-gauge-sec}).
By taking a mass eigenstate basis, we can obtain physical particles, that is,
charged gauge bosons~($W^{\pm}_{\mu}$ and $W^{\prime \pm}_{\mu}$)
and neutral gauge bosons~($Z_{\mu}$, $Z'_{\mu}$ and $A_{\mu}$)
as linear combinations of $W_{0 \mu}$, $W_{1 \mu}$ and $B_{\mu}$.
There are three extra heavy gauge bosons $W^{\prime \pm}$ and $Z'$,
which have almost the same mass.
From experimental constrains on
the $WWZ$ coupling \cite{Hagiwara:1986vm, :2005ema}, 
we get a lower bound on them,
$M_{Z'}\sim M_{W'} \geq 380$ GeV.
Using constraints from $S$ and $T$ parameters,
we can find a upper bound on them,
$M_{Z'}\sim M_{W'} \leq 610$ GeV.

\subsection{Fermion sector}
This model has following fermions;
$\Psi _{L0} \sim (2,1)_{Y}$,
$\Psi _{L1} \sim (1,2)_{Y}$,
$\Psi _{R1} \sim (1,2)_{Y}$
and
$\Psi _{R2} \sim (1,1)_{Q}$,
where
$Y = 1/6$ for quarks
and
$Y = -1/2$ for leptons.
We can write down kinetic terms of fermion by using them. 
The representation of fermions in this model is summarized in Table
\ref{table:fermion_rep}.
\begin{table}[H]
\begin{center}
\begin{tabular}{c|cccc}\hline\hline
            & $SU(2)_0$ & $SU(2)_1$ & $U(1)_2$ & $SU(3)_c$\\ \hline
$\Psi_{L0}$ & $2$  & $1$ & $\frac{1}{6}$
	     $\left(-\frac{1}{2}\right)$ & $3$ ($1$) \\
$\Psi_{L1}$ & $1$ & $2$ & $\frac{1}{6}$
	     $\left(-\frac{1}{2}\right)$ & $3$ ($1$)\\
$\Psi_{R1}$ & $1$ & $2$ & $\frac{1}{6}$
	     $\left(-\frac{1}{2}\right)$ & $3$ ($1$)\\
$\Psi_{R2}
=\left(
\begin{array}{ccc}
u_{R2}  \\
d_{R2}  \\
\end{array} 
\right) $  & $1$ & $1$ & $
\begin{array}{ccc}
\frac{2}{3}  \\
-\frac{1}{3}  \\
\end{array} 
 $ $\left(
\begin{array}{ccc}
0  \\
-1  \\
\end{array} 
\right) $ & $3$ ($1$)\\ 
\hline\hline
\end{tabular} 
\caption{The representation of fermions in this model.
2 and 3 mean fundamental representations for $SU(2)$ and $SU(3)$
 respectively, and 1 means singlet.
 Other numbers are values of hypercharge.
 The numbers in
 parenthesis in columns of $U(1)_2$ and $SU(3)_c$ are for leptons.}
\label{table:fermion_rep}
\end{center}
\end{table}
%
Mass and Yukawa interaction terms in this model are as follows;
\begin{eqnarray*}
  {\cal L}_{\mathrm{fermion}}^{\mathrm{Yukawa}}
=
-
 \sum_{i,j}
&\Biggl[&
  (\overline{\Psi}_{L0})^{i} U _1 m_{1 ij} (\Psi_{R1})^{j}
\\
&&
+ (\overline{\Psi}_{L1})^{i} M_{ij} (\Psi_{R1})^{j} 
\\
&&
+ (\overline{\Psi}_{L1})^{i} U_2  m_{2 ij} (\Psi_{R2})^{j} 
\\
&&
+  (h.c) \frac{}{}
\Biggr],
\label{eq:lagrangian-Yukawa-sec}
\end{eqnarray*}
where
$M$ is Dirac mass,
$i$ and $j$ are indices of generation or flavor,
and
\begin{equation*}
 m_{2 ij}
=
\left(
\begin{array}{cc}
 m_{2u} & 0 \\
 0   & m_{2d} \\
\end{array} 
\right)_{ij}.
\end{equation*}
In general, $m_{1ij}$ and $M_{ij}$ are not flavor blind.
However, to avoid a large FCNC,
$m_{1ij}$ and $M_{ij}$ are assumed to be flavor blind, namely,
$m_{1ij} = m \delta_{ij}$ and $M_{ij} = M \delta_{ij}$.
Under this assumption, the structure of all flavors in this model is embedded in
$m_{2u}$ and $m_{2d}$.
Again, $\Psi$'s are not mass eigenstates but gauge eigenstates.
By diagonalizing mass matrices,
we find that masses of heavy fermions can be described to be approximately $M$.
Using constraints from $S$ and $T$ parameters, 
we can find a lower bound on $M$,
$M \geq 1800$ GeV, which is much heavier than bounds for heavy gauge bosons. 
Therefore the study of production processes of heavy gauge bosons,
is more promissing than heavy fermions at the LHC.

\subsection{Couplings among the heavy gauge boson and light fermions}
\label{sec:constraint_at_tree}

Coupling among $W'$ and light fermions, namely $g_{W'ff}$,
is strongly constrained from electroweak precision measurements \cite{Abe:2008hb}.
Its order of magnitude is 
$g_{W' ff}/g_W \sim {\cal O}(10^{-2})$,
where $g_W = e/s_Z $
and $s_Z^2 \equiv 1 - c_Z^2$,
$c_Z \equiv M_W/M_Z$. 
Coupling among $Z'$ and light fermions are as follows. 
\begin{align*}
 g_{Z' f_L f_L}
&\simeq
g_{W} 
\left(
\frac{\tau^3}{2} G_3
-
t_Z^2 \frac{M_{W}}{M_{W'}}
Q
\right)
,
\\
 g_{Z' f_R f_R}
&\simeq
g_{W} 
\left(
-
t_Z^2 \frac{M_{W}}{M_{W'}}
Q
\right)
,
\end{align*}
where
\begin{align*}
 G_3
&\equiv
\sqrt{2}
\frac{g_{W' ff}}{g_W}
+
t_Z^2 \frac{M_{W}}{M_{W'}}
,
\\
t_Z
&\equiv
\frac{s_Z}{c_Z}
.
\end{align*}
We can see that
$g_{Z' ff}$ is larger than $g_{W'ff}$,
and 
its parameter dependence is moderate compared with $g_{W'ff}$.
Therefore $Z'$ is more suitable than $W'$ as a discovery channel
via DY production process.


\section{Discovery Potential at the LHC}\label{sec:analysis}
In this section, we perform feasibility studies of the Higgsless model at the LHC.
To investigate the $Z'$ discovery potential,
we apply a simple detector simulation with smearing methods,
which approximately reproduce the ATLAS experimental condition at proton-proton collision
of a center-of-mass energy of $\sqrt{s}=14$~TeV~\cite{Aad:2009wy}.
The discovery potential is studied with $Z'\rightarrow WW\rightarrow \ell\nu qq$~($\ell=e$, $\mu$) decay process,
where one of $W$ bosons decays leptonically and the other hadronically.
This channel is capable of reconstructing $Z'$ resonance by solving analytically longitudinal component of a neutrino as described in the following section.

Finally, the discovery potential estimated from invariant mass of two $W$ bosons, $M_{WW}$, as a function of integrated luminosity is shown. 


\subsection{MC sample and cross section}
Signal and dominant background processes are generated with various Monte Carlo~(MC) generators as follows. 
This study uses three preferable $Z'$ signal mass points whose parameters are summarized in Table~\ref{table:sample-points}. 
The $Z'$ signal and $WW$ background processes are generated with CalcHEP~\cite{calchep} and
the parton shower and hadronization are simulated with PYTHIA~\cite{pythia}.
The $t\bar{t}$ process is generated with MC@NLO~\cite{mcatnlo}, $W$+jets with ALPGEN~\cite{alpgen,MLM}
and the parton shower and hadronization are simulated with HERWIG~\cite{herwig}.
The cross section of $Z'$ signal and background processes is summarized in Table~\ref{tbl:Xsec}. 
The ATLAS detector effects are taken into account by
the Monte Carlo simulation smeared with a simplified ATLAS detector~\cite{AcerDET}.

\begin{table}[htbp]
\begin{center}
  \begin{tabular}{ccccc}
   \hline \hline
 $M_{Z'}$(GeV) & $M$ (GeV) & $\frac{g_{W'ff}}{g_W}$ & Br($Z' \to WW$) \\ \hline
  380          & 3500      & 0.023  & 0.971      \\ \hline
  500        & 3500          & 0.023  & 0.987      \\ \hline
  600          & 4300      & 0.022  & 0.991      \\ \hline\hline
 \end{tabular}
\end{center}
\caption{Parameters of $Z'$ signal samples}
\label{table:sample-points}
\end{table}

\begin{table}[htbp]
\begin{center}
\begin{tabular}{cc} \hline\hline
\multicolumn{2}{c}{Signal Sample}\\
$M_{Z'}$ (GeV) & Cross section (pb) \\ \hline
380 & 5.886\\
500 & 4.150\\
600 & 4.165\\ \hline
\multicolumn{2}{c}{Background Sample}\\ 
Process & Cross section (pb) \\ \hline
$WW$ & 111.6~\cite{Aad:2009wy,MCFM} \\
$t\bar{t}$ & 883~\cite{Langenfeld:2009wd} \\
$W(\rightarrow \ell\nu)$+jets & 20510~\cite{Aad:2009wy,PhysRevLett.96.231803} \\
\hline \hline
\end{tabular}
\caption{Cross section of $Z'$ signal and the Standard Model background processes.
The number of $W$+jets is for single lepton flavor.}
\label{tbl:Xsec}
\end{center}
\end{table}


\subsection{Event selection}
The final state of $Z'$ considered in this paper is $\ell\nu qq$.
Experimentally, the neutrino can be observed as a missing transverse energy~(\MET) and
the quark is observed as a jet which is a cluster of hadron.
While the lepton ($e$ and $\mu$) can be measured precisely and used for an event trigger.

First, exactly one high-\pt\ lepton into the detector coverage of a tracking detector ($p_{\mathrm{T}}^{\ell} > 50$ GeV, $|\eta| < 2.5$)\footnote{A pseudo-rapidity $\eta$ is defined by $-2\ln(\tan{\frac{\theta}{2}})$, where $\theta$ is the angle with respect to the beam line.} is required.
The inefficiency of lepton identification and trigger is taken into account and
80\% of efficiency from combined identification and trigger, which is based on MC studies for the ATLAS detector~\cite{Aad:2009wy}, is applied.

Next, a large missing transverse energy (\MET$\ > 50$~GeV) is required.
In addition, exactly two jets with \pt$\ > 50$~GeV and $|\eta| < 3.2$ which is corresponding to the coverage of the calorimeter are required.
If there are jets with \pt$\ > 25$ GeV, $|\eta| < 2.5$ and matched to $b$-quark,
the $b$-tagging which has 50\% efficiency and 2.5$\times 10^{-3}$ false tag rate are applied.
Here we assume that there is no dependence of \pt\ and $\eta$ on the $b$-tagging efficiency and false tag rate.
Events are rejected to reduce enormous $t\bar{t}$ background if at least one $b$-tagged jet exists in the event.
The reconstructed dijet invariant mass, $M_{jj}$, is required to be close to the nominal $W$ mass; 
\[ |M_{jj}-M_{W}| < 15~{\rm GeV}.\]
This selection is very effective to suppress $W(\rightarrow \ell\nu)$+jets background which does not have hadronic decay of $W$ boson. 

Two $W$ bosons decayed from heavy $Z'$ boson are highly boosted.
Hence the events with high $p_{\mathrm{T}}^{\ell\nu}$~(from $p_{\mathrm{T}}^{\ell}$ and \MET) and $p_{\mathrm{T}}^{jj}$ are selected.
The selection criteria depending on $Z'$ mass~($M_{Z'}$) are applied to maximize the discovery potential;
$p_{\mathrm{T}}^{\ell\nu} > 150$, 200 and 250 GeV and $p_{\mathrm{T}}^{jj} > 150$, 200 and 250 GeV are required for $M_{Z'}$ = 380, 500 and 600 GeV, respectively.

$Z'$ invariant mass cannot be reconstructed from observables due to the missing information of
longitudinal neutrino momentum ($p_{z}^{\nu}$).
However we can calculate the longitudinal neutrino momentum by assuming the on-shell $W$ mass constraint\cite{Bach:2009zz};
\begin{equation}\label{eq:WmassConst}
M_{W}^{2} = (E^{\ell}+E^{\nu})^{2}-({\bm p}^{\ell}+{\bm p}^{\nu})^{2},
\end{equation}
where $E^{\ell,\nu}$ and ${\bm p}^{\ell,\nu}$ are energy and momentum of the charged lepton and neutrino, respectively.
The longitudinal component of neutrino momentum is solved analytically from Eq.~(\ref{eq:WmassConst}) as;
\begin{equation}\label{eq:pz}
p_{z}^{\nu} = \frac{(M_{W}^{2}-M_{\ell}^{2}+2{\bm p}_{\mathrm{T}}^{\ell}\cdot{\bm p}_{\mathrm{T}}^{\rm miss})p_{z}^{\ell} \pm \sqrt{D}}{2(E^{\ell 2}-p_{z}^{\ell 2})}
\end{equation}
where ${\bm p}_{\mathrm{T}}^{\ell}$ and ${\bm p}_{\mathrm{T}}^{\rm miss}$ are transverse momentum of charged lepton and neutrino and $D$ is a discriminant represented as 
\begin{eqnarray*}\label{eq:D}
D= E^{\ell 2}\left\{(M_{W}^{2}-M_{\ell}^{2}+2{\bm p}_{\mathrm{T}}^{\ell}\cdot{\bm p}_{\mathrm{T}}^{\rm miss})^{2} \right. \\
\left. -4(E_{\mathrm{T}}^{\rm miss})^{2}(E^{\ell 2}-p_{z}^{\ell 2})\right\}
\end{eqnarray*}
Two solutions of $p_{z}^{\nu}$ can be obtained from Eq.~(\ref{eq:pz}).
It is found that a lower $|p_{z}^{\nu}|$ solution have slightly higher probability to match to the true $p_{z}^{\nu}$ value and
gets better $M_{Z'}$ resolution as shown in Table~\ref{tbl:Solution}.
However a higher $|p_{z}^{\nu}|$ solution also has sufficiently high probability to match to the true $p_{z}^{\nu}$.
In addition, 25\% of events have no solution due to a negative discriminant due to the resolution effect of the smearing.
In this case $D$ is likely close to zero. It is found that $p_{z}^{\nu}$ value corresponding to true value
can be obtained even if the imaginary part is neglected in the $p_{z}^{\nu}$ calculation~($D=0$).

\begin{figure}[htp]
\begin{center}
\includegraphics[width=4cm,height=4cm]{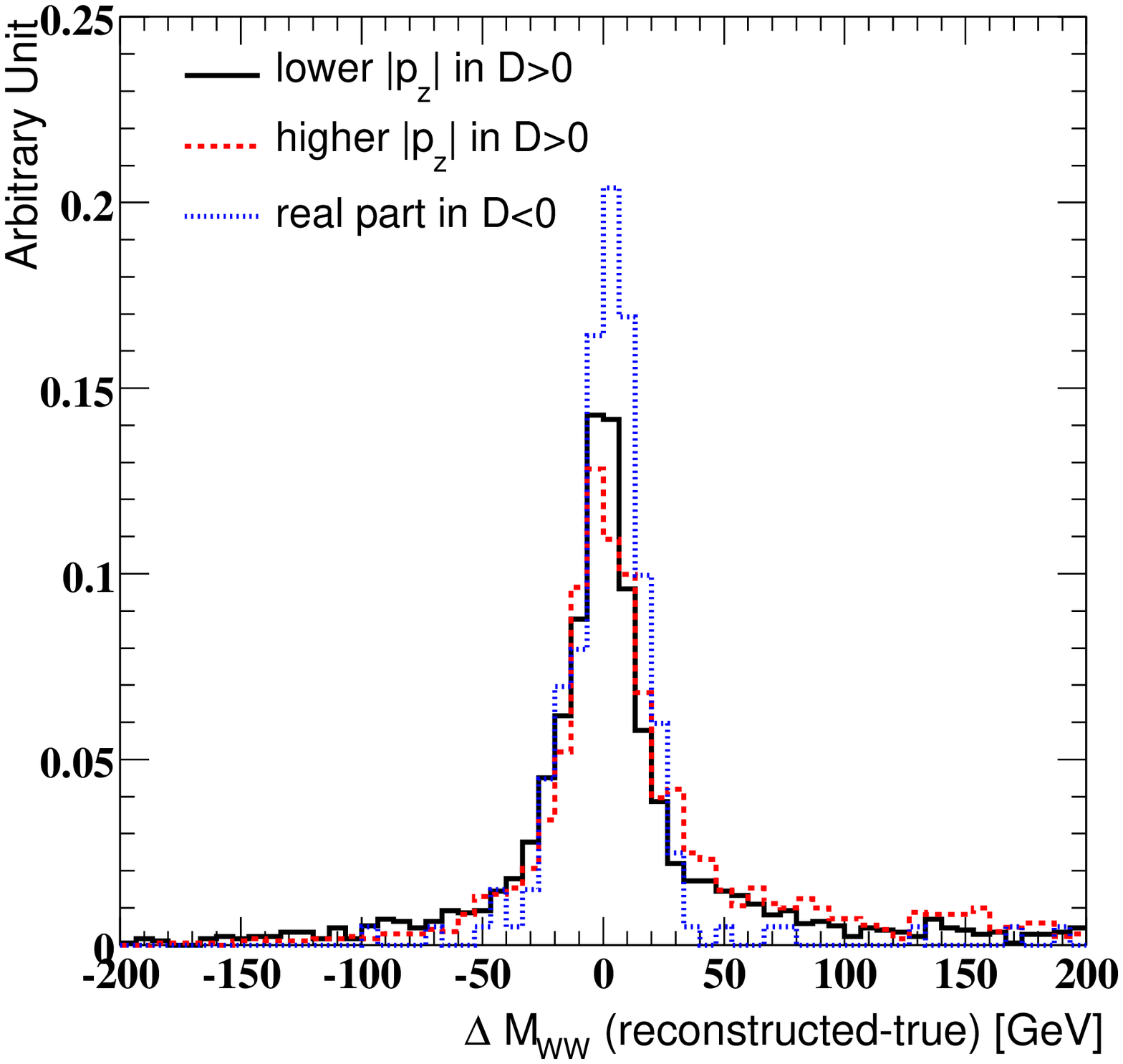}
\includegraphics[width=4cm,height=4cm]{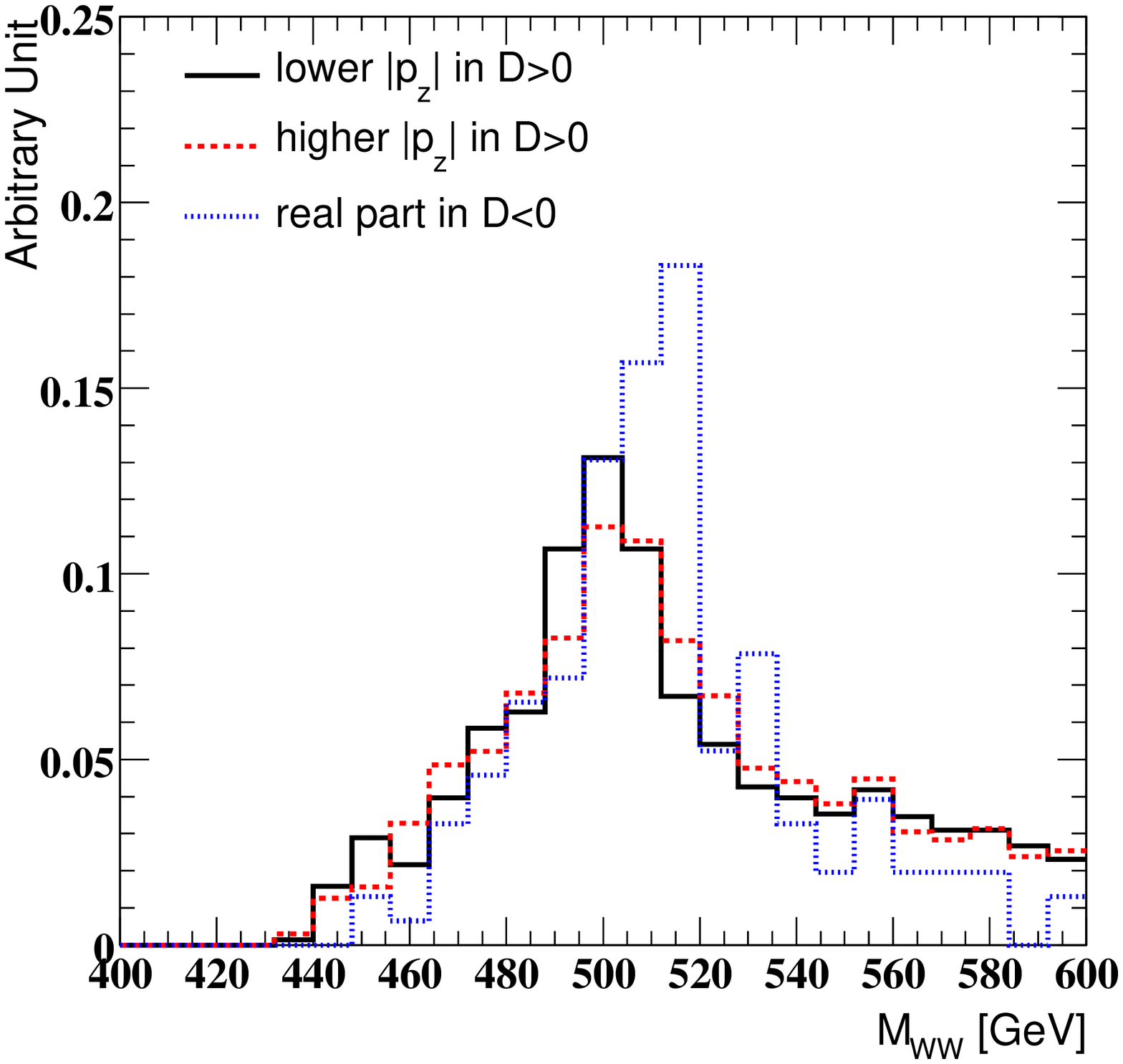}
\caption{Left: $\Delta M_{WW}$(reconstructed-true) for each solution type of longitudinal neutrino momentum. Right: Reconstructed $M_{WW}$ distribution for $M_{Z'}$ = 500 GeV.}
\label{fig:Solution}
\end{center}
\end{figure}

Figure~\ref{fig:Solution}~(left) shows the $\Delta M_{WW}$ distribution for each neutrino solution type in $M_{Z'}$ = 500 GeV,
where $\Delta M_{WW}$ is a difference between a reconstructed $M_{WW}$ and its true value.
Since we adopt all the solutions in any case, 
the correctly reconstructed one has a peak around zero in the $\Delta M_{WW}$ distribution
but the wrongly one makes a tail in the high $\Delta M_{WW}$ region.
This behavior is observed in the reconstructed $M_{Z'}$ distribution as shown in Figure~\ref{fig:Solution}~(right).
We see not only a clear peak but also a long tail in the high mass region.
Table~\ref{tbl:Solution} shows mean and $\sigma$ values for each solution type.
The $M_{WW}$ distribution of a lower $|p_{z}|$ gives the best resolution and
$M_{WW}$ distribution of other solution types can be reconstructed with slightly higher mean value.

In this study, all three solutions are used to maximize a signal acceptance.

\begin{table}[htp]
\begin{center}
\begin{tabular}{crccc}\hline \hline
Solution Type & \multicolumn{2}{c}{Fraction} & Mean (GeV) & $\sigma$ (GeV) \\ \hline
\multirow{2}{*}{Two solutions} &\multirow{2}{*}{75\% $\Bigl\{$}&55\% (lower $|p_{z}|$)& 500.0 & 18.1 \\
                                                      & &45\% (higher $|p_{z}|$)& 503.2 & 21.8 \\
No solution & 25\%&  & 507.2 & 19.6 \\ \hline \hline
\end{tabular}
\caption{The fraction of each solution type to events after event selection and mean and $\sigma$ values extracted from a single Gaussian fit for each solution type.}
\label{tbl:Solution}
\end{center}
\end{table}

\subsection{Discovery potential}
The discovery potential of $Z'$ signal in the three-site Higgsless model is evaluated for three representative mass points.

\begin{figure}[htp]
\begin{center}
\includegraphics[width=7cm,height=6cm]{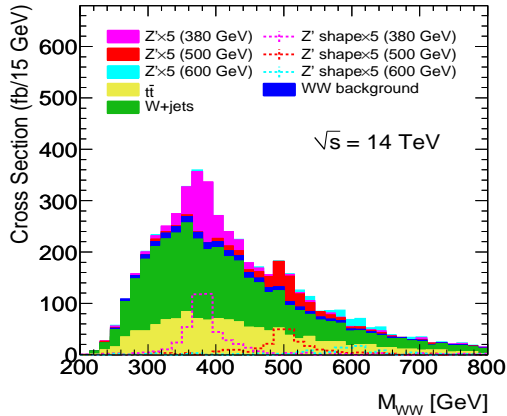}
\caption{Reconstructed $WW$ invariant mass distribution before applying $p_{\mathrm{T}}^{\ell\nu}$, $p_{\mathrm{T}}^{jj}$ selection. All three neutrino solution types are filled in this plot.}
\label{fig:Mww}
\end{center}
\end{figure}

\begin{figure}[htp]
\begin{center}
\includegraphics[width=7cm,height=6cm]{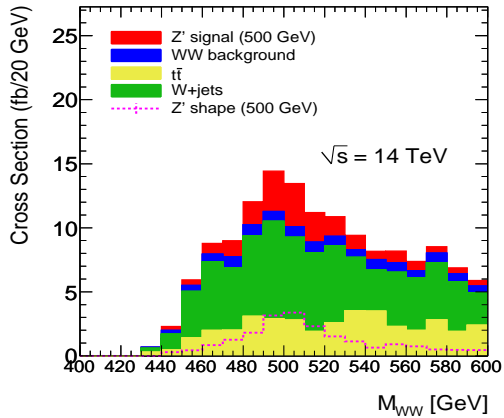}
\caption{Reconstructed $WW$ invariant mass distribution after applying $p_{\mathrm{T}}^{\ell\nu}$, $p_{\mathrm{T}}^{jj}$ selection in $M_{Z'} = 500$ GeV. All three neutrino solution types are filled in this plot.}
\label{fig:Mww500}
\end{center}
\end{figure}

The number of signal events are defined as
\begin{equation*}\label{eq:Nsig}
N_{\rm signal} = N_{{\rm low}\, |p_{z}|}^{2\, {\rm sol}}+N_{{\rm high}\, |p_{z}|}^{2\, {\rm sol}} + N_{\rm real\, part}^{ {\rm No\, sol}},
\end{equation*}
where $N_{{\rm low}\, |p_{z}|}^{2\, {\rm sol}}$ is for a lower $|p_{z}|$ solution,
$N_{{\rm high}\, |p_{z}|}^{2\, {\rm sol}}$ for a higher $|p_{z}|$ and $N_{\rm real\, part}^{ {\rm No\, sol}}$ for the no-solution case.
Similarly, the number of background events~($N_{\rm bkg}$) are defined as the sum of three solutions.
Figure~\ref{fig:Mww} shows the reconstructed $M_{WW}$ distribution before $p_{\mathrm{T}}^{\ell\nu}$ and $p_{\mathrm{T}}^{jj}$ selection. All neutrino solutions are filled in this plot. 
Figure~\ref{fig:Mww500} shows the reconstructed $M_{WW}$ distribution after $p_{\mathrm{T}}^{\ell\nu}$ and $p_{\mathrm{T}}^{jj}$ selection in $M_{Z'}$ = 500 GeV. 

The discovery potential is evaluated from the number of expected signal and background in a $M_{Z'}$ mass window, which is determined to get the maximum significance.
This significance is defined as follows;
\begin{equation*}\label{eq:Significance}
{\rm Significance} = \frac{N_{\rm signal}}{\sqrt{N_{\rm bkg}}}\;.
\end{equation*}

Table~\ref{tbl:FinalNumber} shows the number of expected signal and background into the signal mass window for each signal mass point.
The bottom line shows significance. The number of events are normalized to 1~\fb.
\begin{table}[htp]
\begin{center}
\begin{tabular}{cccc}\hline \hline
\multicolumn{4}{c}{Expected cross section with 1~\fb ($\sqrt{s} = 14$~TeV)} \\ \hline
$M_{Z'}$ (GeV)    & 380   & 500   & 600 \\\hline
Optimized Mass window (GeV) & 360-400 & 460-540 & 580-660 \\ \hline
$W$+jets          & 53.7  & 45.9  & 9.25  \\
$t\bar{t}$        & 19.8  & 19.9  & 4.75  \\
$WW$              & 5.73  & 5.91  & 1.48  \\ 
Total background  & 79.2  & 71.7  & 15.5 \\ \hline
Signal            & 25.5  & 15.3  & 4.6  \\ \hline
Significance      & 2.86  & 1.81  & 1.17\\ \hline \hline
\end{tabular}
\caption{The number of signal and background in the signal mass window region for each mass point and obtained significance at 1~\fb. These numbers include all three solutions.}
\label{tbl:FinalNumber}
\end{center}
\end{table}

\begin{figure}[htp]
\begin{center}
\includegraphics[width=7cm,height=6cm]{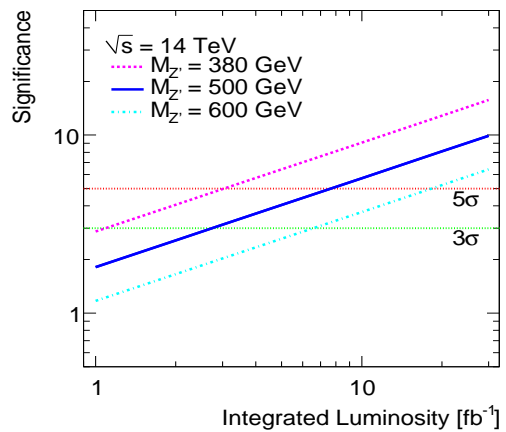}
\caption{Discovery potential for each $M_{Z'}$ sample as a function of integrated luminosity. Two vertical lines correspond to the 3$\sigma$ evidence and 5$\sigma$ discovery thresholds, respectively.}
\label{fig:FinalResult}
\end{center}
\end{figure}

Figure~\ref{fig:FinalResult} shows the expected significance as a function of integrated luminosity for each mass point.
The significance reaches 3$\sigma$ threshold for $M_{Z'}$ = 380, 500 and 600 GeV in about 1~\fb, 2.5~\fb\ and 6~\fb, and 5$\sigma$ threshold for $M_{Z'}$ = 380, 500 and 600 GeV in about 3~\fb, 8~\fb\ and 20~\fb, respectively. 

\section{Summary and Discussion}\label{sec:summary}
We have studied the discovery potential of the three-site Higgsless model
with the $Z'$ gauge boson via Drell-Yan production at proton-proton collision of a center-of-mass energy of $\sqrt{s}$ = 14~TeV. 
The discovery potential of $Z'\rightarrow WW\rightarrow \ell\nu qq$ is evaluated with the simplified detector simulation of ATLAS experiment condition
which takes into account the effect of hadronization and experimental efficiency and resolution.
The significance is obtained with the event counting in the signal mass window.
We show that the significance reaches 3$\sigma$ threshold in the integrated luminosity 1-6 \fb~
and 5 $\sigma$ threshold in the 3-20 \fb~ for theoretically preferable $Z'$ mass region, 380-610 GeV.
The required integrated luminosity for the observation corresponds to a few years' running of the LHC at $\sqrt{s}$ = 14~TeV.

%
%
Depending on the model parameters,
we have found less integrated luminosity is required for the discovery
of the $Z'$ boson through the Drell-Yan process than other production processes.
On the other hand, 
for the discovery of the $W^{\prime \pm}$ boson, 
other production processes have advantage rather than the
Drell-Yan process
because of the fermiophobity of the $W^{\prime \pm}$ boson.
It is important to discover the $W^{\prime \pm}$ boson  
because we have to check the origin of the $Z'$ boson is the $SU(2)$
gauge symmetry for a verification of Higgsless models.
There are many models which predict 
the $Z'$ boson
which is not assosiated with $SU(2)$ gauge symmetry\cite{Langacker:2008yv}. 
Precision predictions in such $Z'$ production are studied
in\cite{Fuks:2007gk,Fuks:2008mi,Rizzo:2006nw,Dittmar:2003ir}, for example.  
Hence we need to discover $W^{\prime \pm}$ for a verification of Higgsless
models.
There is a mutually complementary relationship between the Drell-Yan
process and other production processes
in order to verify Higgsless models at the LHC.

\newpage
\section*{Acknowledgments}
T.A. is supported in part by the JSPS Grant-in-Aid No. 204354.


\bibliography{lhc3site}

\begin{thebibliography}{60}
\expandafter\ifx\csname natexlab\endcsname\relax\def\natexlab#1{#1}\fi
\expandafter\ifx\csname bibnamefont\endcsname\relax
  \def\bibnamefont#1{#1}\fi
\expandafter\ifx\csname bibfnamefont\endcsname\relax
  \def\bibfnamefont#1{#1}\fi
\expandafter\ifx\csname citenamefont\endcsname\relax
  \def\citenamefont#1{#1}\fi
\expandafter\ifx\csname url\endcsname\relax
  \def\url#1{\texttt{#1}}\fi
\expandafter\ifx\csname urlprefix\endcsname\relax\def\urlprefix{URL }\fi
\providecommand{\bibinfo}[2]{#2}
\providecommand{\eprint}[2][]{\url{#2}}

\bibitem[{\citenamefont{Evans and Bryant}(2008)}]{LHC}
\bibinfo{author}{\bibfnamefont{L.}~\bibnamefont{Evans}} \bibnamefont{and}
  \bibinfo{author}{\bibfnamefont{P.}~\bibnamefont{Bryant}},
  \bibinfo{journal}{JINST} \textbf{\bibinfo{volume}{3}},
  \bibinfo{pages}{S08001} (\bibinfo{year}{2008}).

\bibitem[{\citenamefont{Herrero}(1998)}]{Herrero:1998eq}
\bibinfo{author}{\bibfnamefont{M.}~\bibnamefont{Herrero}}
  (\bibinfo{year}{1998}), \bibinfo{note}{{\em Proc. 10th NATO ASI on Techniques
  and Concepts of High-Energy Physics, St. Croix, Virgin Islands, U.S.A., June
  18--29, 1998}}, \eprint{hep-ph/9812242}.

\bibitem[{\citenamefont{Dicus and Mathur}(1973)}]{Dicus:1992vj}
\bibinfo{author}{\bibfnamefont{D.~A.} \bibnamefont{Dicus}} \bibnamefont{and}
  \bibinfo{author}{\bibfnamefont{V.~S.} \bibnamefont{Mathur}},
  \bibinfo{journal}{Phys. Rev.} \textbf{\bibinfo{volume}{D7}},
  \bibinfo{pages}{3111} (\bibinfo{year}{1973}).

\bibitem[{\citenamefont{Cornwall et~al.}(1973)\citenamefont{Cornwall, Levin,
  and Tiktopoulos}}]{Cornwall:1973tb}
\bibinfo{author}{\bibfnamefont{J.~M.} \bibnamefont{Cornwall}},
  \bibinfo{author}{\bibfnamefont{D.~N.} \bibnamefont{Levin}}, \bibnamefont{and}
  \bibinfo{author}{\bibfnamefont{G.}~\bibnamefont{Tiktopoulos}},
  \bibinfo{journal}{Phys. Rev. Lett.} \textbf{\bibinfo{volume}{30}},
  \bibinfo{pages}{1268} (\bibinfo{year}{1973}).

\bibitem[{\citenamefont{Cornwall et~al.}(1974)\citenamefont{Cornwall, Levin,
  and Tiktopoulos}}]{Cornwall:1974km}
\bibinfo{author}{\bibfnamefont{J.~M.} \bibnamefont{Cornwall}},
  \bibinfo{author}{\bibfnamefont{D.~N.} \bibnamefont{Levin}}, \bibnamefont{and}
  \bibinfo{author}{\bibfnamefont{G.}~\bibnamefont{Tiktopoulos}},
  \bibinfo{journal}{Phys. Rev.} \textbf{\bibinfo{volume}{D10}},
  \bibinfo{pages}{1145} (\bibinfo{year}{1974}).

\bibitem[{\citenamefont{Lee et~al.}(1977{\natexlab{a}})\citenamefont{Lee,
  Quigg, and Thacker}}]{Lee:1977yc}
\bibinfo{author}{\bibfnamefont{B.~W.} \bibnamefont{Lee}},
  \bibinfo{author}{\bibfnamefont{C.}~\bibnamefont{Quigg}}, \bibnamefont{and}
  \bibinfo{author}{\bibfnamefont{H.~B.} \bibnamefont{Thacker}},
  \bibinfo{journal}{Phys. Rev. Lett.} \textbf{\bibinfo{volume}{38}},
  \bibinfo{pages}{883} (\bibinfo{year}{1977}{\natexlab{a}}).

\bibitem[{\citenamefont{Lee et~al.}(1977{\natexlab{b}})\citenamefont{Lee,
  Quigg, and Thacker}}]{Lee:1977eg}
\bibinfo{author}{\bibfnamefont{B.~W.} \bibnamefont{Lee}},
  \bibinfo{author}{\bibfnamefont{C.}~\bibnamefont{Quigg}}, \bibnamefont{and}
  \bibinfo{author}{\bibfnamefont{H.~B.} \bibnamefont{Thacker}},
  \bibinfo{journal}{Phys. Rev.} \textbf{\bibinfo{volume}{D16}},
  \bibinfo{pages}{1519} (\bibinfo{year}{1977}{\natexlab{b}}).

\bibitem[{\citenamefont{Chanowitz and Gaillard}(1985)}]{Chanowitz:1985hj}
\bibinfo{author}{\bibfnamefont{M.~S.} \bibnamefont{Chanowitz}}
  \bibnamefont{and} \bibinfo{author}{\bibfnamefont{M.~K.}
  \bibnamefont{Gaillard}}, \bibinfo{journal}{Nucl. Phys.}
  \textbf{\bibinfo{volume}{B261}}, \bibinfo{pages}{379} (\bibinfo{year}{1985}).

\bibitem[{\citenamefont{Csaki et~al.}(2004{\natexlab{a}})\citenamefont{Csaki,
  Grojean, Murayama, Pilo, and Terning}}]{Csaki:2003dt}
\bibinfo{author}{\bibfnamefont{C.}~\bibnamefont{Csaki}},
  \bibinfo{author}{\bibfnamefont{C.}~\bibnamefont{Grojean}},
  \bibinfo{author}{\bibfnamefont{H.}~\bibnamefont{Murayama}},
  \bibinfo{author}{\bibfnamefont{L.}~\bibnamefont{Pilo}}, \bibnamefont{and}
  \bibinfo{author}{\bibfnamefont{J.}~\bibnamefont{Terning}},
  \bibinfo{journal}{Phys. Rev.} \textbf{\bibinfo{volume}{D69}},
  \bibinfo{pages}{055006} (\bibinfo{year}{2004}{\natexlab{a}}),
  \eprint{hep-ph/0305237}.

\bibitem[{\citenamefont{Csaki et~al.}(2004{\natexlab{b}})\citenamefont{Csaki,
  Grojean, Pilo, and Terning}}]{Csaki:2003zu}
\bibinfo{author}{\bibfnamefont{C.}~\bibnamefont{Csaki}},
  \bibinfo{author}{\bibfnamefont{C.}~\bibnamefont{Grojean}},
  \bibinfo{author}{\bibfnamefont{L.}~\bibnamefont{Pilo}}, \bibnamefont{and}
  \bibinfo{author}{\bibfnamefont{J.}~\bibnamefont{Terning}},
  \bibinfo{journal}{Phys. Rev. Lett.} \textbf{\bibinfo{volume}{92}},
  \bibinfo{pages}{101802} (\bibinfo{year}{2004}{\natexlab{b}}),
  \eprint{hep-ph/0308038}.

\bibitem[{\citenamefont{Nomura}(2003)}]{Nomura:2003du}
\bibinfo{author}{\bibfnamefont{Y.}~\bibnamefont{Nomura}},
  \bibinfo{journal}{JHEP} \textbf{\bibinfo{volume}{11}}, \bibinfo{pages}{050}
  (\bibinfo{year}{2003}), \eprint{hep-ph/0309189}.

\bibitem[{\citenamefont{Barbieri et~al.}(2004)\citenamefont{Barbieri, Pomarol,
  and Rattazzi}}]{Barbieri:2003pr}
\bibinfo{author}{\bibfnamefont{R.}~\bibnamefont{Barbieri}},
  \bibinfo{author}{\bibfnamefont{A.}~\bibnamefont{Pomarol}}, \bibnamefont{and}
  \bibinfo{author}{\bibfnamefont{R.}~\bibnamefont{Rattazzi}},
  \bibinfo{journal}{Phys. Lett.} \textbf{\bibinfo{volume}{B591}},
  \bibinfo{pages}{141} (\bibinfo{year}{2004}), \eprint{hep-ph/0310285}.

\bibitem[{\citenamefont{Csaki et~al.}(2004{\natexlab{c}})\citenamefont{Csaki,
  Grojean, Hubisz, Shirman, and Terning}}]{Csaki:2003sh}
\bibinfo{author}{\bibfnamefont{C.}~\bibnamefont{Csaki}},
  \bibinfo{author}{\bibfnamefont{C.}~\bibnamefont{Grojean}},
  \bibinfo{author}{\bibfnamefont{J.}~\bibnamefont{Hubisz}},
  \bibinfo{author}{\bibfnamefont{Y.}~\bibnamefont{Shirman}}, \bibnamefont{and}
  \bibinfo{author}{\bibfnamefont{J.}~\bibnamefont{Terning}},
  \bibinfo{journal}{Phys. Rev.} \textbf{\bibinfo{volume}{D70}},
  \bibinfo{pages}{015012} (\bibinfo{year}{2004}{\natexlab{c}}),
  \eprint{hep-ph/0310355}.

\bibitem[{\citenamefont{Davoudiasl et~al.}(2004)\citenamefont{Davoudiasl,
  Hewett, Lillie, and Rizzo}}]{Davoudiasl:2003me}
\bibinfo{author}{\bibfnamefont{H.}~\bibnamefont{Davoudiasl}},
  \bibinfo{author}{\bibfnamefont{J.~L.} \bibnamefont{Hewett}},
  \bibinfo{author}{\bibfnamefont{B.}~\bibnamefont{Lillie}}, \bibnamefont{and}
  \bibinfo{author}{\bibfnamefont{T.~G.} \bibnamefont{Rizzo}},
  \bibinfo{journal}{Phys. Rev.} \textbf{\bibinfo{volume}{D70}},
  \bibinfo{pages}{015006} (\bibinfo{year}{2004}), \eprint{hep-ph/0312193}.

\bibitem[{\citenamefont{Burdman and Nomura}(2004)}]{Burdman:2003ya}
\bibinfo{author}{\bibfnamefont{G.}~\bibnamefont{Burdman}} \bibnamefont{and}
  \bibinfo{author}{\bibfnamefont{Y.}~\bibnamefont{Nomura}},
  \bibinfo{journal}{Phys. Rev.} \textbf{\bibinfo{volume}{D69}},
  \bibinfo{pages}{115013} (\bibinfo{year}{2004}), \eprint{hep-ph/0312247}.

\bibitem[{\citenamefont{Foadi et~al.}(2004)\citenamefont{Foadi, Gopalakrishna,
  and Schmidt}}]{Foadi:2003xa}
\bibinfo{author}{\bibfnamefont{R.}~\bibnamefont{Foadi}},
  \bibinfo{author}{\bibfnamefont{S.}~\bibnamefont{Gopalakrishna}},
  \bibnamefont{and} \bibinfo{author}{\bibfnamefont{C.}~\bibnamefont{Schmidt}},
  \bibinfo{journal}{JHEP} \textbf{\bibinfo{volume}{03}}, \bibinfo{pages}{042}
  (\bibinfo{year}{2004}), \eprint{hep-ph/0312324}.

\bibitem[{\citenamefont{Hirn and Stern}(2004)}]{Hirn:2004ze}
\bibinfo{author}{\bibfnamefont{J.}~\bibnamefont{Hirn}} \bibnamefont{and}
  \bibinfo{author}{\bibfnamefont{J.}~\bibnamefont{Stern}},
  \bibinfo{journal}{Eur. Phys. J.} \textbf{\bibinfo{volume}{C34}},
  \bibinfo{pages}{447} (\bibinfo{year}{2004}), \eprint{hep-ph/0401032}.

\bibitem[{\citenamefont{Casalbuoni et~al.}(2004)\citenamefont{Casalbuoni,
  De~Curtis, and Dominici}}]{Casalbuoni:2004id}
\bibinfo{author}{\bibfnamefont{R.}~\bibnamefont{Casalbuoni}},
  \bibinfo{author}{\bibfnamefont{S.}~\bibnamefont{De~Curtis}},
  \bibnamefont{and} \bibinfo{author}{\bibfnamefont{D.}~\bibnamefont{Dominici}},
  \bibinfo{journal}{Phys. Rev.} \textbf{\bibinfo{volume}{D70}},
  \bibinfo{pages}{055010} (\bibinfo{year}{2004}), \eprint{hep-ph/0405188}.

\bibitem[{\citenamefont{Chivukula et~al.}(2004)\citenamefont{Chivukula,
  Simmons, He, Kurachi, and Tanabashi}}]{Chivukula:2004pk}
\bibinfo{author}{\bibfnamefont{R.~S.} \bibnamefont{Chivukula}},
  \bibinfo{author}{\bibfnamefont{E.~H.} \bibnamefont{Simmons}},
  \bibinfo{author}{\bibfnamefont{H.-J.} \bibnamefont{He}},
  \bibinfo{author}{\bibfnamefont{M.}~\bibnamefont{Kurachi}}, \bibnamefont{and}
  \bibinfo{author}{\bibfnamefont{M.}~\bibnamefont{Tanabashi}},
  \bibinfo{journal}{Phys. Rev.} \textbf{\bibinfo{volume}{D70}},
  \bibinfo{pages}{075008} (\bibinfo{year}{2004}), \eprint{hep-ph/0406077}.

\bibitem[{\citenamefont{Csaki et~al.}(2005)\citenamefont{Csaki, Hubisz, and
  Meade}}]{Csaki:2005vy}
\bibinfo{author}{\bibfnamefont{C.}~\bibnamefont{Csaki}},
  \bibinfo{author}{\bibfnamefont{J.}~\bibnamefont{Hubisz}}, \bibnamefont{and}
  \bibinfo{author}{\bibfnamefont{P.}~\bibnamefont{Meade}}
  (\bibinfo{year}{2005}), \bibinfo{note}{{\em Proc. TASI2004: Physics in $D\ge
  4$ Boulder, Colorado, U.S.A., June 6 -- July 2, 2004}},
  \eprint{hep-ph/0510275}.

\bibitem[{\citenamefont{Csaki}(2004)}]{Csaki:2004sz}
\bibinfo{author}{\bibfnamefont{C.}~\bibnamefont{Csaki}} (\bibinfo{year}{2004}),
  \bibinfo{note}{{\em Proc. 12th International Conference on Supersymmetry and
  Unification of Fundamental Interactions (SUSU04), Tsukuba, Japan, June
  17--23, 2004}}, \eprint{hep-ph/0412339}.

\bibitem[{\citenamefont{Arkani-Hamed et~al.}(2001)\citenamefont{Arkani-Hamed,
  Cohen, and Georgi}}]{Arkanihamed:2001ca}
\bibinfo{author}{\bibfnamefont{N.}~\bibnamefont{Arkani-Hamed}},
  \bibinfo{author}{\bibfnamefont{A.~G.} \bibnamefont{Cohen}}, \bibnamefont{and}
  \bibinfo{author}{\bibfnamefont{H.}~\bibnamefont{Georgi}},
  \bibinfo{journal}{Phys. Rev. Lett.} \textbf{\bibinfo{volume}{86}},
  \bibinfo{pages}{4757} (\bibinfo{year}{2001}), \eprint{hep-th/0104005}.

\bibitem[{\citenamefont{Hill et~al.}(2001)\citenamefont{Hill, Pokorski, and
  Wang}}]{Hill:2000mu}
\bibinfo{author}{\bibfnamefont{C.~T.} \bibnamefont{Hill}},
  \bibinfo{author}{\bibfnamefont{S.}~\bibnamefont{Pokorski}}, \bibnamefont{and}
  \bibinfo{author}{\bibfnamefont{J.}~\bibnamefont{Wang}},
  \bibinfo{journal}{Phys. Rev.} \textbf{\bibinfo{volume}{D64}},
  \bibinfo{pages}{105005} (\bibinfo{year}{2001}), \eprint{hep-th/0104035}.

\bibitem[{\citenamefont{Georgi}(1989)}]{Georgi:1989gp}
\bibinfo{author}{\bibfnamefont{H.}~\bibnamefont{Georgi}},
  \bibinfo{journal}{Phys. Rev. Lett.} \textbf{\bibinfo{volume}{63}},
  \bibinfo{pages}{1917} (\bibinfo{year}{1989}).

\bibitem[{\citenamefont{Georgi}(1990)}]{Georgi:1989xy}
\bibinfo{author}{\bibfnamefont{H.}~\bibnamefont{Georgi}},
  \bibinfo{journal}{Nucl. Phys.} \textbf{\bibinfo{volume}{B331}},
  \bibinfo{pages}{311} (\bibinfo{year}{1990}).

\bibitem[{\citenamefont{Bando et~al.}(1985{\natexlab{a}})\citenamefont{Bando,
  Kugo, Uehara, Yamawaki, and Yanagida}}]{Bando:1984ej}
\bibinfo{author}{\bibfnamefont{M.}~\bibnamefont{Bando}},
  \bibinfo{author}{\bibfnamefont{T.}~\bibnamefont{Kugo}},
  \bibinfo{author}{\bibfnamefont{S.}~\bibnamefont{Uehara}},
  \bibinfo{author}{\bibfnamefont{K.}~\bibnamefont{Yamawaki}}, \bibnamefont{and}
  \bibinfo{author}{\bibfnamefont{T.}~\bibnamefont{Yanagida}},
  \bibinfo{journal}{Phys. Rev. Lett.} \textbf{\bibinfo{volume}{54}},
  \bibinfo{pages}{1215} (\bibinfo{year}{1985}{\natexlab{a}}).

\bibitem[{\citenamefont{Bando et~al.}(1985{\natexlab{b}})\citenamefont{Bando,
  Kugo, and Yamawaki}}]{Bando:1985rf}
\bibinfo{author}{\bibfnamefont{M.}~\bibnamefont{Bando}},
  \bibinfo{author}{\bibfnamefont{T.}~\bibnamefont{Kugo}}, \bibnamefont{and}
  \bibinfo{author}{\bibfnamefont{K.}~\bibnamefont{Yamawaki}},
  \bibinfo{journal}{Nucl. Phys.} \textbf{\bibinfo{volume}{B259}},
  \bibinfo{pages}{493} (\bibinfo{year}{1985}{\natexlab{b}}).

\bibitem[{\citenamefont{Bando et~al.}(1988{\natexlab{a}})\citenamefont{Bando,
  Fujiwara, and Yamawaki}}]{Bando:1987ym}
\bibinfo{author}{\bibfnamefont{M.}~\bibnamefont{Bando}},
  \bibinfo{author}{\bibfnamefont{T.}~\bibnamefont{Fujiwara}}, \bibnamefont{and}
  \bibinfo{author}{\bibfnamefont{K.}~\bibnamefont{Yamawaki}},
  \bibinfo{journal}{Prog. Theor. Phys.} \textbf{\bibinfo{volume}{79}},
  \bibinfo{pages}{1140} (\bibinfo{year}{1988}{\natexlab{a}}).

\bibitem[{\citenamefont{Bando et~al.}(1988{\natexlab{b}})\citenamefont{Bando,
  Kugo, and Yamawaki}}]{Bando:1987br}
\bibinfo{author}{\bibfnamefont{M.}~\bibnamefont{Bando}},
  \bibinfo{author}{\bibfnamefont{T.}~\bibnamefont{Kugo}}, \bibnamefont{and}
  \bibinfo{author}{\bibfnamefont{K.}~\bibnamefont{Yamawaki}},
  \bibinfo{journal}{Phys. Rept.} \textbf{\bibinfo{volume}{164}},
  \bibinfo{pages}{217} (\bibinfo{year}{1988}{\natexlab{b}}).

\bibitem[{\citenamefont{Harada and Yamawaki}(2003)}]{Harada:2003jx}
\bibinfo{author}{\bibfnamefont{M.}~\bibnamefont{Harada}} \bibnamefont{and}
  \bibinfo{author}{\bibfnamefont{K.}~\bibnamefont{Yamawaki}},
  \bibinfo{journal}{Phys. Rept.} \textbf{\bibinfo{volume}{381}},
  \bibinfo{pages}{1} (\bibinfo{year}{2003}), \eprint{hep-ph/0302103}.

\bibitem[{\citenamefont{S.~Chivukula et~al.}(2006)\citenamefont{S.~Chivukula,
  Coleppa, Di~Chiara, H.~Simmons, He, Kurachi, and
  Tanabashi}}]{SekharChivukula:2006cg}
\bibinfo{author}{\bibfnamefont{R.}~\bibnamefont{S.~Chivukula}},
  \bibinfo{author}{\bibfnamefont{B.}~\bibnamefont{Coleppa}},
  \bibinfo{author}{\bibfnamefont{S.}~\bibnamefont{Di~Chiara}},
  \bibinfo{author}{\bibfnamefont{E.}~\bibnamefont{H.~Simmons}},
  \bibinfo{author}{\bibfnamefont{H.}~\bibnamefont{He}},
  \bibinfo{author}{\bibfnamefont{M.}~\bibnamefont{Kurachi}}, \bibnamefont{and}
  \bibinfo{author}{\bibfnamefont{M.}~\bibnamefont{Tanabashi}},
  \bibinfo{journal}{Phys. Rev.} \textbf{\bibinfo{volume}{D74}},
  \bibinfo{pages}{075011} (\bibinfo{year}{2006}), \eprint{hep-ph/0607124}.

\bibitem[{\citenamefont{Abe et~al.}(2008)\citenamefont{Abe, Matsuzaki, and
  Tanabashi}}]{Abe:2008hb}
\bibinfo{author}{\bibfnamefont{T.}~\bibnamefont{Abe}},
  \bibinfo{author}{\bibfnamefont{S.}~\bibnamefont{Matsuzaki}},
  \bibnamefont{and}
  \bibinfo{author}{\bibfnamefont{M.}~\bibnamefont{Tanabashi}},
  \bibinfo{journal}{Phys. Rev.} \textbf{\bibinfo{volume}{D78}},
  \bibinfo{pages}{055020} (\bibinfo{year}{2008}), \eprint{0807.2298}.

\bibitem[{\citenamefont{Abe et~al.}(2009)}]{Abe:2009ni}
\bibinfo{author}{\bibfnamefont{T.}~\bibnamefont{Abe}} \bibnamefont{et~al.},
  \bibinfo{journal}{Phys. Rev.} \textbf{\bibinfo{volume}{D79}},
  \bibinfo{pages}{075016} (\bibinfo{year}{2009}), \eprint{0902.3910}.

\bibitem[{\citenamefont{Ohl and Speckner}(2008)}]{Ohl:2008ri}
\bibinfo{author}{\bibfnamefont{T.}~\bibnamefont{Ohl}} \bibnamefont{and}
  \bibinfo{author}{\bibfnamefont{C.}~\bibnamefont{Speckner}},
  \bibinfo{journal}{Phys. Rev.} \textbf{\bibinfo{volume}{D78}},
  \bibinfo{pages}{095008} (\bibinfo{year}{2008}), \eprint{0809.0023}.

\bibitem[{\citenamefont{Speckner}(2010)}]{Speckner:2010zi}
\bibinfo{author}{\bibfnamefont{C.}~\bibnamefont{Speckner}}
  (\bibinfo{year}{2010}), \eprint{1011.1851}.

\bibitem[{\citenamefont{Han et~al.}(2010)\citenamefont{Han, Krohn, Wang, and
  Zhu}}]{Han:2009em}
\bibinfo{author}{\bibfnamefont{T.}~\bibnamefont{Han}},
  \bibinfo{author}{\bibfnamefont{D.}~\bibnamefont{Krohn}},
  \bibinfo{author}{\bibfnamefont{L.-T.} \bibnamefont{Wang}}, \bibnamefont{and}
  \bibinfo{author}{\bibfnamefont{W.}~\bibnamefont{Zhu}},
  \bibinfo{journal}{JHEP} \textbf{\bibinfo{volume}{03}}, \bibinfo{pages}{082}
  (\bibinfo{year}{2010}), \eprint{0911.3656}.

\bibitem[{\citenamefont{Asano and Shimizu}(2010)}]{Asano:2010ii}
\bibinfo{author}{\bibfnamefont{M.}~\bibnamefont{Asano}} \bibnamefont{and}
  \bibinfo{author}{\bibfnamefont{Y.}~\bibnamefont{Shimizu}}
  (\bibinfo{year}{2010}), \eprint{1010.5230}.

\bibitem[{\citenamefont{Han et~al.}(2009)\citenamefont{Han, Liu, Luo, Wang, and
  Wu}}]{Han:2009qr}
\bibinfo{author}{\bibfnamefont{T.}~\bibnamefont{Han}},
  \bibinfo{author}{\bibfnamefont{H.-S.} \bibnamefont{Liu}},
  \bibinfo{author}{\bibfnamefont{M.-x.} \bibnamefont{Luo}},
  \bibinfo{author}{\bibfnamefont{K.}~\bibnamefont{Wang}}, \bibnamefont{and}
  \bibinfo{author}{\bibfnamefont{W.}~\bibnamefont{Wu}}, \bibinfo{journal}{Phys.
  Rev.} \textbf{\bibinfo{volume}{D80}}, \bibinfo{pages}{095010}
  (\bibinfo{year}{2009}), \eprint{0908.2186}.

\bibitem[{\citenamefont{Bian et~al.}(2009)}]{Bian:2009kf}
\bibinfo{author}{\bibfnamefont{J.-G.} \bibnamefont{Bian}} \bibnamefont{et~al.},
  \bibinfo{journal}{Nucl. Phys.} \textbf{\bibinfo{volume}{B819}},
  \bibinfo{pages}{201} (\bibinfo{year}{2009}), \eprint{0905.2336}.

\bibitem[{\citenamefont{He et~al.}(2008)}]{He:2007ge}
\bibinfo{author}{\bibfnamefont{H.-J.} \bibnamefont{He}} \bibnamefont{et~al.},
  \bibinfo{journal}{Phys. Rev.} \textbf{\bibinfo{volume}{D78}},
  \bibinfo{pages}{031701} (\bibinfo{year}{2008}), \eprint{0708.2588}.

\bibitem[{\citenamefont{Bach}(2009)}]{Bach:2009zz}
\bibinfo{author}{\bibfnamefont{F.}~\bibnamefont{Bach}} (\bibinfo{year}{2009}),
  \bibinfo{note}{phenomenology of the Three Site Higgsless Model at the ATLAS
  Detector of the LHC, {\em Master's Thesis}}.

\bibitem[{\citenamefont{Aad et~al.}(2008)}]{ATLASDetector}
\bibinfo{author}{\bibfnamefont{G.}~\bibnamefont{Aad}} \bibnamefont{et~al.}
  (\bibinfo{collaboration}{The ATLAS Collaboration}), \bibinfo{journal}{JINST}
  \textbf{\bibinfo{volume}{3}}, \bibinfo{pages}{S08003} (\bibinfo{year}{2008}).

\bibitem[{\citenamefont{Hagiwara et~al.}(1987)\citenamefont{Hagiwara, Peccei,
  Zeppenfeld, and Hikasa}}]{Hagiwara:1986vm}
\bibinfo{author}{\bibfnamefont{K.}~\bibnamefont{Hagiwara}},
  \bibinfo{author}{\bibfnamefont{R.~D.} \bibnamefont{Peccei}},
  \bibinfo{author}{\bibfnamefont{D.}~\bibnamefont{Zeppenfeld}},
  \bibnamefont{and} \bibinfo{author}{\bibfnamefont{K.}~\bibnamefont{Hikasa}},
  \bibinfo{journal}{Nucl. Phys.} \textbf{\bibinfo{volume}{B282}},
  \bibinfo{pages}{253} (\bibinfo{year}{1987}).

\bibitem[{:20(2006)}]{:2005ema}
\bibinfo{journal}{Phys.Rept.} \textbf{\bibinfo{volume}{427}},
  \bibinfo{pages}{257} (\bibinfo{year}{2006}), \eprint{hep-ex/0509008}.

\bibitem[{\citenamefont{Aad et~al.}(2009)}]{Aad:2009wy}
\bibinfo{author}{\bibfnamefont{G.}~\bibnamefont{Aad}} \bibnamefont{et~al.}
  (\bibinfo{collaboration}{The ATLAS Collaboration}) (\bibinfo{year}{2009}),
  \eprint{0901.0512}.

\bibitem[{\citenamefont{Alexander~Pukhov}()}]{calchep}
\bibinfo{author}{\bibfnamefont{N.~C.} \bibnamefont{Alexander~Pukhov},
  \bibfnamefont{Alexander~Belyaev}},
  \bibinfo{howpublished}{http://theory.sinp.msu.ru/\~{}pukhov/calchep.html}.

\bibitem[{\citenamefont{Sjostrand et~al.}(2006)\citenamefont{Sjostrand, Mrenna,
  and Skands}}]{pythia}
\bibinfo{author}{\bibfnamefont{T.}~\bibnamefont{Sjostrand}},
  \bibinfo{author}{\bibfnamefont{S.}~\bibnamefont{Mrenna}}, \bibnamefont{and}
  \bibinfo{author}{\bibfnamefont{P.~Z.} \bibnamefont{Skands}},
  \bibinfo{journal}{JHEP} \textbf{\bibinfo{volume}{05}}, \bibinfo{pages}{026}
  (\bibinfo{year}{2006}).

\bibitem[{\citenamefont{Frixione and Webber}(2003)}]{mcatnlo}
\bibinfo{author}{\bibfnamefont{S.}~\bibnamefont{Frixione}} \bibnamefont{and}
  \bibinfo{author}{\bibfnamefont{B.}~\bibnamefont{Webber}},
  \bibinfo{journal}{JHEP} \textbf{\bibinfo{volume}{0308}}, \bibinfo{pages}{007}
  (\bibinfo{year}{2003}).

\bibitem[{\citenamefont{Mangano et~al.}(2003)}]{alpgen}
\bibinfo{author}{\bibfnamefont{M.~L.} \bibnamefont{Mangano}}
  \bibnamefont{et~al.}, \bibinfo{journal}{JHEP} \textbf{\bibinfo{volume}{07}},
  \bibinfo{pages}{001} (\bibinfo{year}{2003}).

\bibitem[{\citenamefont{Hoche et~al.}(2006)}]{MLM}
\bibinfo{author}{\bibfnamefont{S.}~\bibnamefont{Hoche}} \bibnamefont{et~al.},
  \bibinfo{journal}{hep-ph/0602031}  (\bibinfo{year}{2006}).

\bibitem[{\citenamefont{Corcella et~al.}(2001)}]{herwig}
\bibinfo{author}{\bibfnamefont{G.}~\bibnamefont{Corcella}}
  \bibnamefont{et~al.}, \bibinfo{journal}{JHEP} \textbf{\bibinfo{volume}{01}},
  \bibinfo{pages}{010} (\bibinfo{year}{2001}).

\bibitem[{\citenamefont{Richter-Was}(2002)}]{AcerDET}
\bibinfo{author}{\bibfnamefont{E.}~\bibnamefont{Richter-Was}},
  \bibinfo{journal}{hep-ph/0207355}  (\bibinfo{year}{2002}).

\bibitem[{\citenamefont{J and K}()}]{MCFM}
\bibinfo{author}{\bibfnamefont{C.}~\bibnamefont{J}} \bibnamefont{and}
  \bibinfo{author}{\bibfnamefont{E.}~\bibnamefont{K}},
  \bibinfo{howpublished}{http://mcfm.fnal.gov/}.

\bibitem[{\citenamefont{Langenfeld et~al.}(2009)\citenamefont{Langenfeld, Moch,
  and Uwer}}]{Langenfeld:2009wd}
\bibinfo{author}{\bibfnamefont{U.}~\bibnamefont{Langenfeld}},
  \bibinfo{author}{\bibfnamefont{S.}~\bibnamefont{Moch}}, \bibnamefont{and}
  \bibinfo{author}{\bibfnamefont{P.}~\bibnamefont{Uwer}},
  \bibinfo{journal}{Phys. Rev.} \textbf{\bibinfo{volume}{D80}},
  \bibinfo{pages}{054009} (\bibinfo{year}{2009}), \eprint{0906.5273}.

\bibitem[{\citenamefont{Melnikov and Petriello}(2006)}]{PhysRevLett.96.231803}
\bibinfo{author}{\bibfnamefont{K.}~\bibnamefont{Melnikov}} \bibnamefont{and}
  \bibinfo{author}{\bibfnamefont{F.}~\bibnamefont{Petriello}},
  \bibinfo{journal}{Phys. Rev. Lett.} \textbf{\bibinfo{volume}{96}},
  \bibinfo{pages}{231803} (\bibinfo{year}{2006}).

\bibitem[{\citenamefont{Langacker}(2009)}]{Langacker:2008yv}
\bibinfo{author}{\bibfnamefont{P.}~\bibnamefont{Langacker}},
  \bibinfo{journal}{Rev.Mod.Phys.} \textbf{\bibinfo{volume}{81}},
  \bibinfo{pages}{1199} (\bibinfo{year}{2009}), \eprint{0801.1345}.

\bibitem[{\citenamefont{Fuks et~al.}(2008{\natexlab{a}})\citenamefont{Fuks,
  Klasen, Ledroit, Li, and Morel}}]{Fuks:2007gk}
\bibinfo{author}{\bibfnamefont{B.}~\bibnamefont{Fuks}},
  \bibinfo{author}{\bibfnamefont{M.}~\bibnamefont{Klasen}},
  \bibinfo{author}{\bibfnamefont{F.}~\bibnamefont{Ledroit}},
  \bibinfo{author}{\bibfnamefont{Q.}~\bibnamefont{Li}}, \bibnamefont{and}
  \bibinfo{author}{\bibfnamefont{J.}~\bibnamefont{Morel}},
  \bibinfo{journal}{Nucl.Phys.} \textbf{\bibinfo{volume}{B797}},
  \bibinfo{pages}{322} (\bibinfo{year}{2008}{\natexlab{a}}),
  \eprint{0711.0749}.

\bibitem[{\citenamefont{Fuks et~al.}(2008{\natexlab{b}})\citenamefont{Fuks,
  van~der Bij, and Xu}}]{Fuks:2008mi}
\bibinfo{author}{\bibfnamefont{B.}~\bibnamefont{Fuks}},
  \bibinfo{author}{\bibfnamefont{J.~J.} \bibnamefont{van~der Bij}},
  \bibnamefont{and} \bibinfo{author}{\bibfnamefont{Q.}~\bibnamefont{Xu}},
  \bibinfo{journal}{Phys.Rev.} \textbf{\bibinfo{volume}{D78}},
  \bibinfo{pages}{074016} (\bibinfo{year}{2008}{\natexlab{b}}),
  \eprint{0808.1680}.

\bibitem[{\citenamefont{Rizzo}(2006)}]{Rizzo:2006nw}
\bibinfo{author}{\bibfnamefont{T.~G.} \bibnamefont{Rizzo}}, pp.
  \bibinfo{pages}{537--575} (\bibinfo{year}{2006}), \eprint{hep-ph/0610104}.

\bibitem[{\citenamefont{Dittmar et~al.}(2004)\citenamefont{Dittmar, Nicollerat,
  and Djouadi}}]{Dittmar:2003ir}
\bibinfo{author}{\bibfnamefont{M.}~\bibnamefont{Dittmar}},
  \bibinfo{author}{\bibfnamefont{A.-S.} \bibnamefont{Nicollerat}},
  \bibnamefont{and} \bibinfo{author}{\bibfnamefont{A.}~\bibnamefont{Djouadi}},
  \bibinfo{journal}{Phys.Lett.} \textbf{\bibinfo{volume}{B583}},
  \bibinfo{pages}{111} (\bibinfo{year}{2004}), \eprint{hep-ph/0307020}.

\end{thebibliography}
\end{document}